# The long-term solar variability, as reconstructed from historical sources: Several case studies in the 17th – 18th centuries

Hisashi Hayakawa (1 – 4)*

(1) Institute for Space-Earth Environmental Research, Nagofya University, Nagoya, Japan
(2) Institute for Advanced Research, Nagoya University, Nagoya, Japan
(3) Space Physics and Operations Division, RAL Space, Science and Technology Facilities Council, Rutherford Appleton Laboratory, Harwell Oxford, Didcot, UK
(4) Astro-Glaciology Laboratory, Riken Nishina Centre, Wako, Japan

* hisashi@nagoya-u.jp

## Abstract

On a centennial timescale, solar activity was quantified based on records of instrumental sunspot observations. This article briefly discusses several aspects of the recent archival investigations of historical sunspot records in the 17th to 18th centuries. This article also reviews the recent updates for the active day fraction and positions of the reported sunspot groups of the Maunder Minimum to show their significance within the observational history. These archival investigations serve as base datasets for reconstructing solar activity.

## 1. Introduction

The Sun is one of the most frequently observed fixed stars by the human beings. Solar activity has been quantified using multiple observational indices, such as solar radio emissions at multiple wavelengths and sunspot datasets (Hathaway, 2015). The relative sunspot and sunspot group numbers are unique indices for quantitatively covering solar activity on a centennial timescale (Clette *et al*., 2023; Usoskin, 2023). These indices have been reconstructed from individual instrumental sunspot observations since 1610 (Hoyt and Schatten, 1998; Vaquero *et al*., 2016; Clette *et al*., 2023).

Following Hoyt and Schatten (1998), these indices were developed and recalibrated using multiple methods (Clette and Lefèvre, 2016; Svalgaard and Schatten, 2016; Usoskin *et al*., 2016, 2021a; Chatzistergos *et al*., 2017). However, their reconstructions show substantial discrepancies before 1900, as shown in figure 2 of Muñoz-Jaramillo and Vaquero (2019). This emphasizes the importance of improving calibration methods and source datasets. Several studies have been published since Hoyt and Schatten (1998), which have been summarized by Vaquero *et al*. (2016). Eight years have passed since then and these studies have made considerable progress. This article briefly discusses several aspects of the recent developments.

**2. The Early 17th Century**

This was where instrumental sunspot observations started. At the time of writing, the earliest achievements in telescopic sunspot observations were associated with Harriot (18 Dec 1610) in terms of data observations and with Fabricius (Jun 1611) in terms of publication (Vaquero *et al*., 2016; Arlt and Vaquero, 2020). Numerous early astronomers followed these early observations, as exemplified with Galilei, Scheiner, Mögling, Gassendi, and Hevelius (Figure 1). These records were subsequently exploited to derive sunspot group numbers and positions (Arlt *et al*., 2016; Vokhmyanin and Zolotova, 2018a, 2018b; Carrasco *et al*., 2019; Hayakawa *et al*., 2021a).

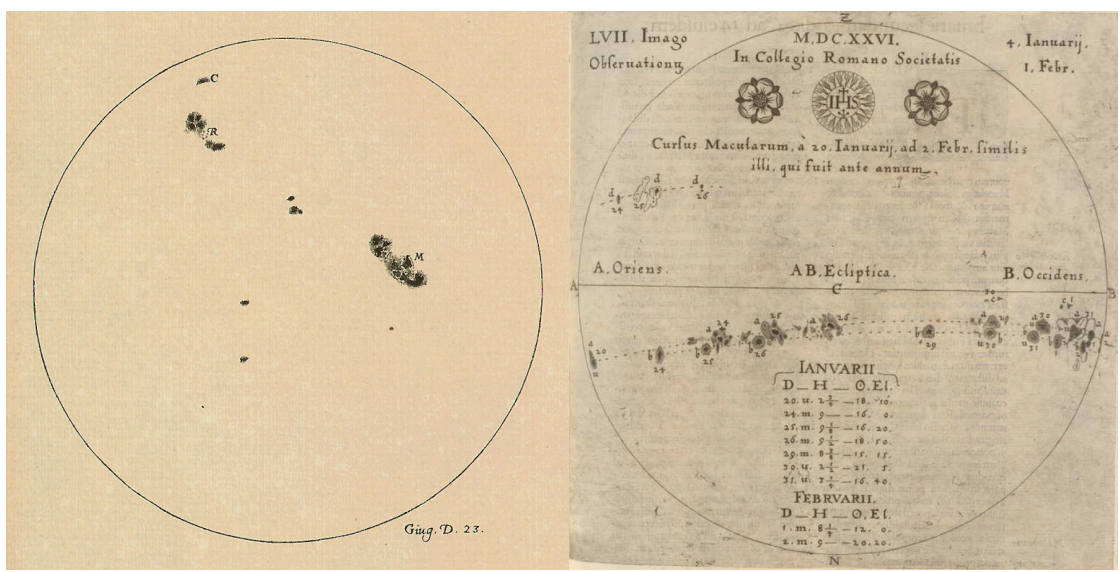

Figure 1: Examples of early sunspot drawings. Left panel shows Galilei's sunspot drawing in 1612 Jun (Galilei, 1613, p. 79). Right panel shows Scheiner's sunspot drawing in 1626 Jan – Feb (Scheiner, 1630, p. 297).

These records capture three solar cycles from the earliest telescopic sunspot record in 1610 to the beginning of the Maunder Minimum (1645 – 1715). These sunspot records captured the first solar cycle incompletely, as this solar cycle seemed to have started before 1610. However, determining when the solar cycle began is difficult. Cosmogenic isotope data have been used to indirectly reconstruct solar cycles over longer time scales, including at the beginning of the 17th century. However, these studies have shown conflicting reconstructions for this solar cycle, lasting approximately 16 years from 1606 to 1622 (Miyahara *et al*., 2021) and approximately 11 years from 1609 to 1620 (Usoskin *et al*., 2021b), as shown in Figure 2. Notably, resolving this discrepancy is important because the anomalous duration of this solar cycle has been associated with the precursor of the Maunder Minimum, a unique grand minimum of solar activity in the last four centuries (Miyahara *et al*., 2021).

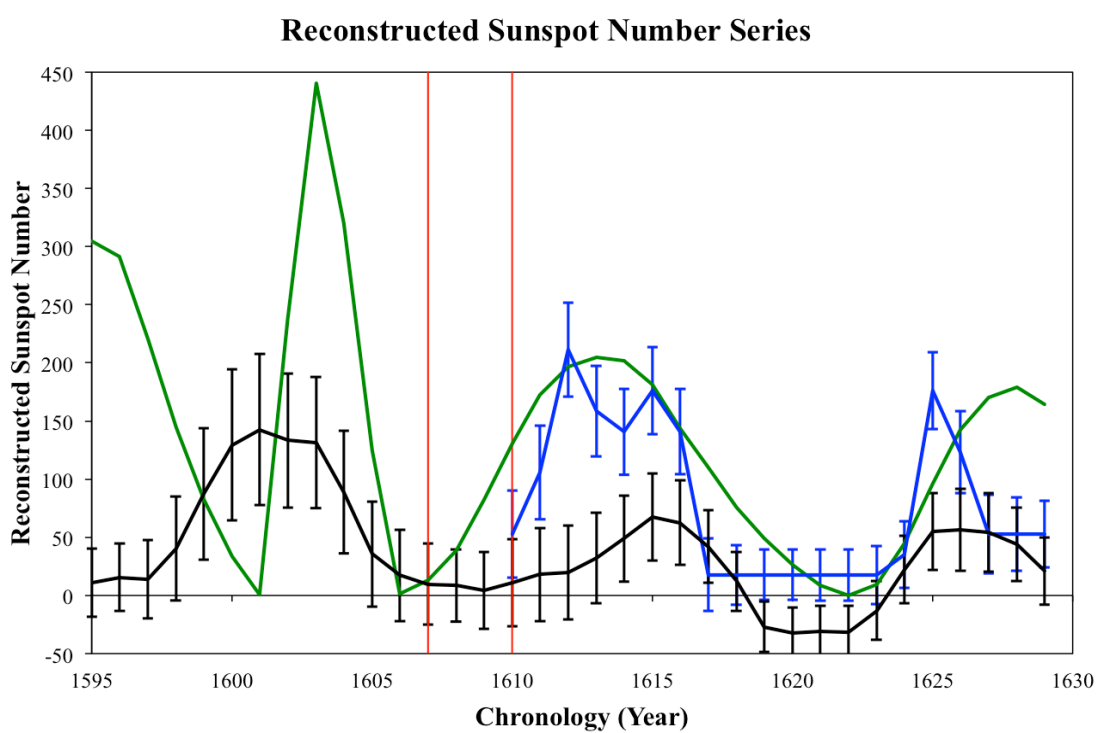

Figure 2: Hayakawa *et al*.'s solar-cycle minimum constraints for the first telescopic solar cycle in the 1610s are shown in red lines as their boundaries. These constraints are compared with reconstructed sunspot number based on $^{14}$C data in a black curve (Usoskin *et al*., 2021b) and a green curve (Miyahara *et al*., 2021), as well as sunspot group number of Svalgaard and Schatten (2016), multiplied by factor of 20, according to Muñoz-Jaramillo and Vaquero (2019), shown in a blue curve.

This figure is reproduced from Hayakawa *et al*. (2024b).

In this context, recollecting the history of astronomy at the time was crucial. During this period, telescopes were not the only instruments used for solar observation. Some observers also used cameras that were obscured. Kepler observed a solar disk in search of Mercury crossing it (Kepler, 1609). Kepler's report recorded a naked-eye sunspot group, not Mercury crossing the solar disk, on 28 May 1607 with two whole-disk drawings (Figure 3). As shown in Hayakawa *et al*. (2024b), Kepler's records allowed us to locate this sunspot group between S17° and N12° in heliographic latitude (Figure 4).

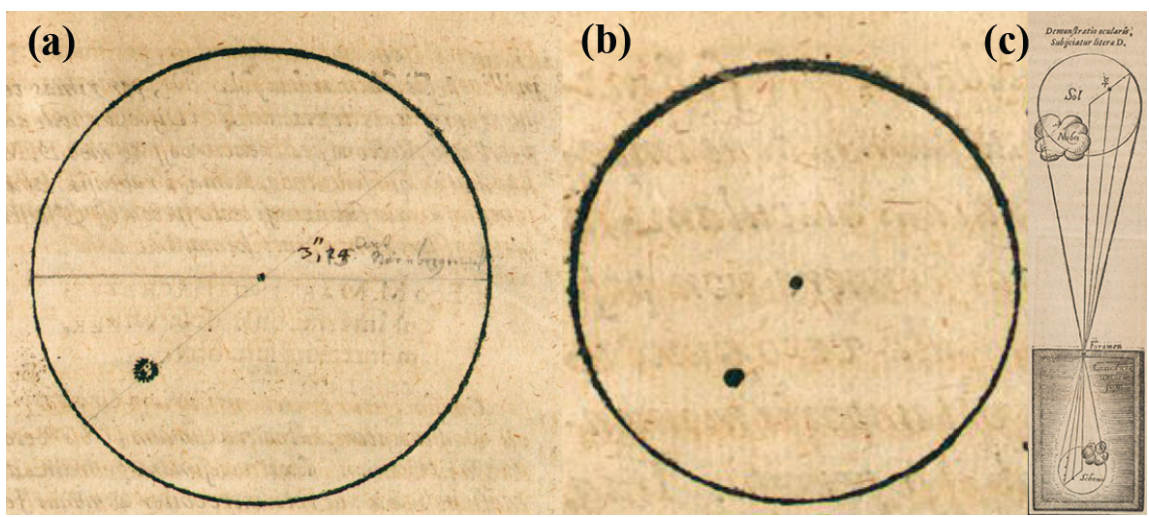

Figure 3: Kepler's sunspot drawings in Kepler (1609), as reproduced from Hayakawa *et al*. (2024b). Following Spörer's law (Section 4.10 of Hathaway (2015)), modern statistics show a much higher probability of locating this to the tail of the previous solar cycle (≈ 80%) than to the beginning of the new solar cycle (≈ 20%). This is consistent with the contrast of early telescopic sunspot records showing sunspot positions at higher heliographic latitudes (Figure 4).

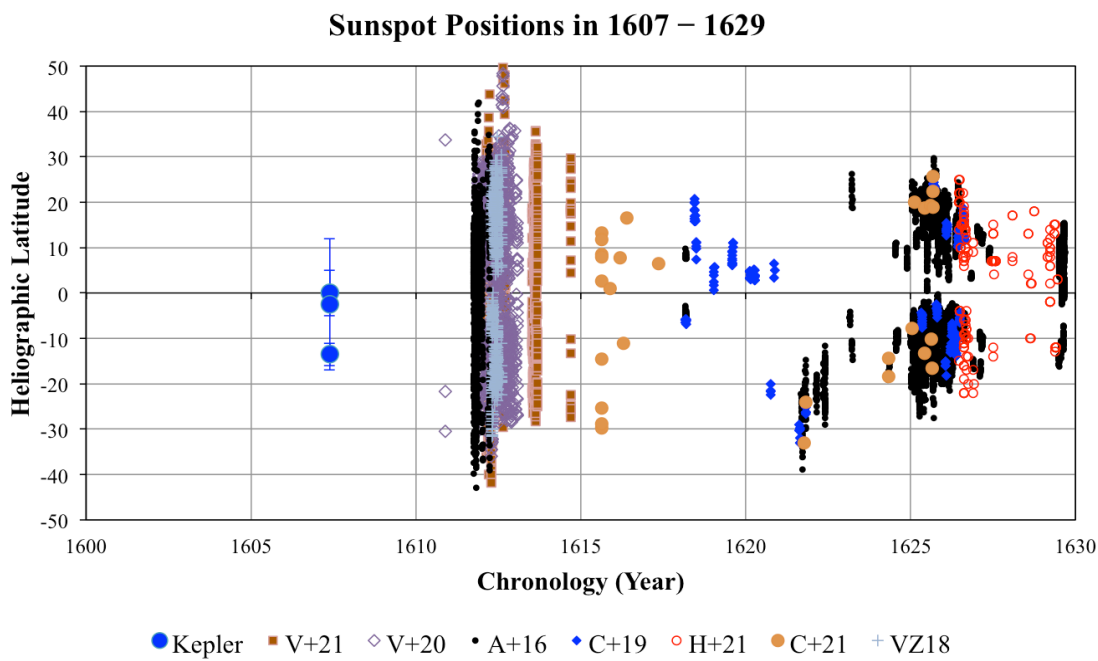

Figure 4: Distributions of the reported sunspot latitudes in 1607 – 1629, contrasting our three scenarios for Kepler (Hayakawa *et al*., 2024b), in comparison with V+21 (Vokhmyanin *et al*., 2021), V+20 (Vokhmyanin *et al*., 2020), A+16 (Arlt *et al*., 2016), C+19 (Carrasco *et al*., 2019), H+21 (Hayakawa *et al*., 2021b), C+21 (Carrasco *et al*., 2021a), and VZ18 (Vokhmyanin and Zolotova, 2018a). This figure is reproduced from Hayakawa *et al*. (2024b)

Kepler's records contributed to modern science as one of the earliest graphical records of instrumental sunspot observations, predating Harriot's sunspot records. Kepler's records allowed us to locate the onset of the first telescopic solar cycle between 1607 and 1610 (Figure 2). This is consistent with Usoskin *et al*.'s reconstruction, which contradicts that of Miyahara *et al*. (2021). The first telescopic solar cycle was probably not extremely long, but likely had a normal duration of ≈ 11 years.

## 3. The Maunder Minimum

The Maunder Minimum is a unique period with extremely small solar cycles, during which most of the number of the sunspot group was extremely low (Usoskin *et al*., 2015; Vaquero *et al*., 2015a; Carrasco *et al*., 2022), the reported sunspot groups appear in the southern solar hemisphere (Ribes and Nesme-Ribes, 1993) and solar coronal streamers are missing in visual accounts (Eddy, 1976; Riley *et al*., 2015; Hayakawa *et al*., 2021b). This seems to have been a unique period in the last four centuries, as shown in figure 2 of Muñoz-Jaramillo and Vaquero (2019).

However, we need to be cautious about the reliability of existing datasets, as there were considerable contaminations, such as those from observations for different purposes and general descriptions. These cases require serious revisions and reanalyses for the source records in this period. There are some cases where records are added from previously forgotten records. As such, Hoyt and Schatten's data have been subjected to fundamental revisions. The revised database was released as Vaquero *et al*. (2016). Afterward, several studies have been published for sunspot records in the Maunder Minimum. These studies have revised records for the Eimmart Observatory, J. H. Müller, Hoffmann, and Wideburg in Hayakawa *et al*. (2021d), Fogelius, Siverus, Picard, Kircher, and Hook in Hayakawa *et al*. (2021c), Kirch Family, Ihle, Schultz, Strum, and Hertel in Neuhäuser *et al*. (2018), and early sunspot records in 1645 – 1659 in Hayakawa *et al.* (2024a). Additionally, Carrasco *et al*. (2021b) added records of Gallet in 1677. These updates are mostly reviewed in Section 2.2.1(b) of Clette *et al*. (2023).

This article subsequently shows some of their cases. Hoyt and Schatten (1998) cited Keill as claiming that there were continuous spotless days between May 1653 and June 1670. However, Keill's original statements appear considerably different. Keill stated: '... but (*mais*) from 1653 until (*jusqu'en*) 1670 at best one or two were discovered; since then, they have reappeared quite often in abundance. Apparently, they do not follow any rule in their appearance [Original Text: ... ***mais depuis 1653 jusqu'en 1670** à peine en a-t-on découvert une ou deux; depuis elles ont reparu assez souvent en abondance. Il semble qu'elles ne suivent aucune loi dans leur apparition*]' (Keill, 1746, p. 52; emphases added). Keill does not mention the months. They are likely a misinterpretation of two words: *mais* vs *mai* (but vs May) and *jusqu'en* vs *juin* (until vs Jun).

Similar cases were found in the sunspot records of Fogelius and Siverus species. Hoyt and Schatten (1998) associated these observers with 3605 and 5400 days of sunspot observation. Vaquero *et al*. (2016) raised doubts regarding the reliability of these continuous spotless days, although their source records were not exploited. Their source reports show that the majority of Fogelius' records come from his quote from Picard mentioning his previous sunspot observations in 1661. These were not Fogelius's experiences. Moreover, what has been considered as Fogelius' sunspot drawing has been confirmed as Siverus' sunspot drawing, based on his signature of the original manuscript. As such, archival investigations by Hayakawa *et al*. (2021c) reduced the number of Fogelius and Siverus from

92

3605 days and 5400 days to 3 and 25 days, respectively. They formed warning examples of contamination based on the general descriptions.

Similarly, we must be cautious about the contamination of different types of observations in the sunspot database. For example, Mouton's reports (Mouton, 1670) were associated with considerable spotless days in Hoyt and Schatten (1998) and Vaquero *et al*. (2016) with a considerable number of spotless days. However, his report only discusses his measurements of solar diameter from 1659 to 1661. Mouton (1670) did not document the presence or absence of sunspots. This is also the case with Hevelius' reports from 1652 to 1659, in which we found contamination from solar altitude measurements.

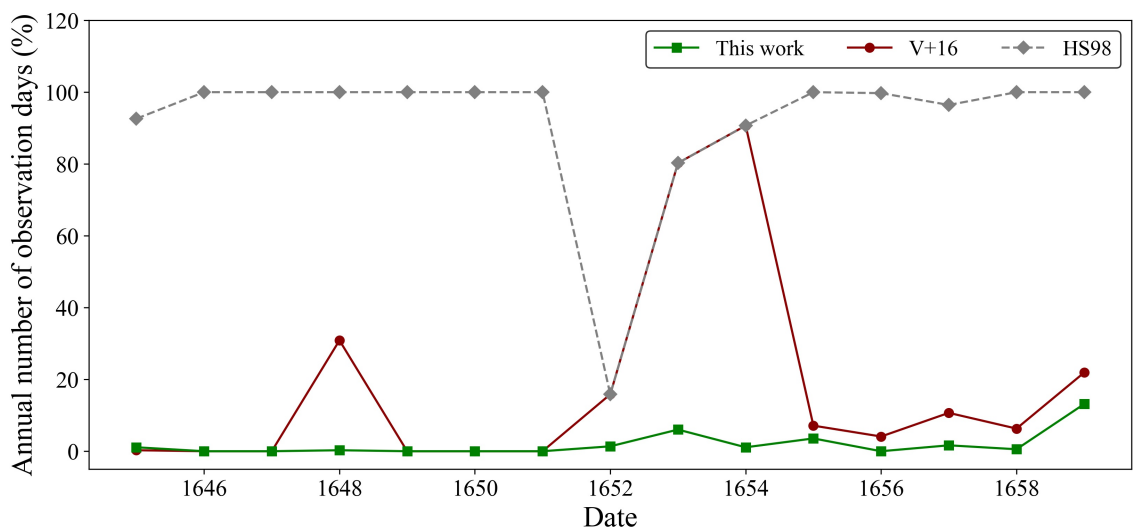

Figure 5: Percentage of the annual number of days with observations in the early Maunder Minimum, according to Hoyt and Schatten (1998) noted with grey diamonds (HS98), Vaquero *et al*. (2016) noted with red circles (V+16) and Hayakawa *et al*. (2024a) noted with green squares (this study), as reproduced from Hayakawa *et al*. (2024a).

These cases substantially reduced the amount of data in the sunspot dataset. Hayakawa *et al*. (2024a) showed the actual coverage of sunspot records in 1645 – 1659 was approximately 2%, while Hoyt and Schatten (1998) and Vaquero *et al*. (2016) considered approximately 92% and 18% of the dates in 1645 – 1659 was with sunspot records (Figure 5).

These modifications inevitably change the active day fraction (ADF), which is the ratio of observational dates with and without sunspots. The ADF has a good correlation with the sunspot number, especially when solar activity is sufficiently low; hence, it has been used to estimate the solar activity level in the Maunder Minimum (Vaquero *et al*., 2015). Subsequently, Carrasco *et al*. (2022) compiled revised datasets that are available at that time, calculated the ADF and sunspot number in the Maunder Minimum, and compared their reconstruction with Usoskin *et al*.'s sunspot number reconstruction based on the $^{14}$C data of tree ring archives. Hayakawa *et al*. (2024b) have reviewed and revised all the sunspot records in 1645 – 1659, calculated the ADF and sunspot number in the said period, and compared their reconstruction with Usoskin *et al*.'s sunspot number reconstruction.

Figure 6 shows the ADF data based on all the data that are currently available at the time of writing. They are generally the same with Carrasco *et al*. (2022) with updates of Hayakawa *et al*. (2024b) without using new data, while some miscalculations have been corrected here (*e.g.*, the ADF value in 1652). Caveats must be noted here, as they are subject to further revisions and probably showing lower value than the reality owing to numerous contaminations to be removed. In any case, these values show substantially lower ADF than normal solar cycles or even their minima (*e.g.*, Carrasco *et al*., 2021c).

These values were partially used to calculate the relative number of sunspots. This result is consistent with that of Usoskin *et al*. (2021b), who reconstructed $^{14}$C data from tree rings (figure 5 of Carrasco *et al*. (2022) and figure 5 of Hayakawa *et al*. (2024a)). Similar analyses were performed for the aftermath of the Maunder Minimum using the records of Rost and Alischer (Carrasco *et al*., 2024). They confirmed gradual transitions between the Maunder Minimum and normal solar cycles.

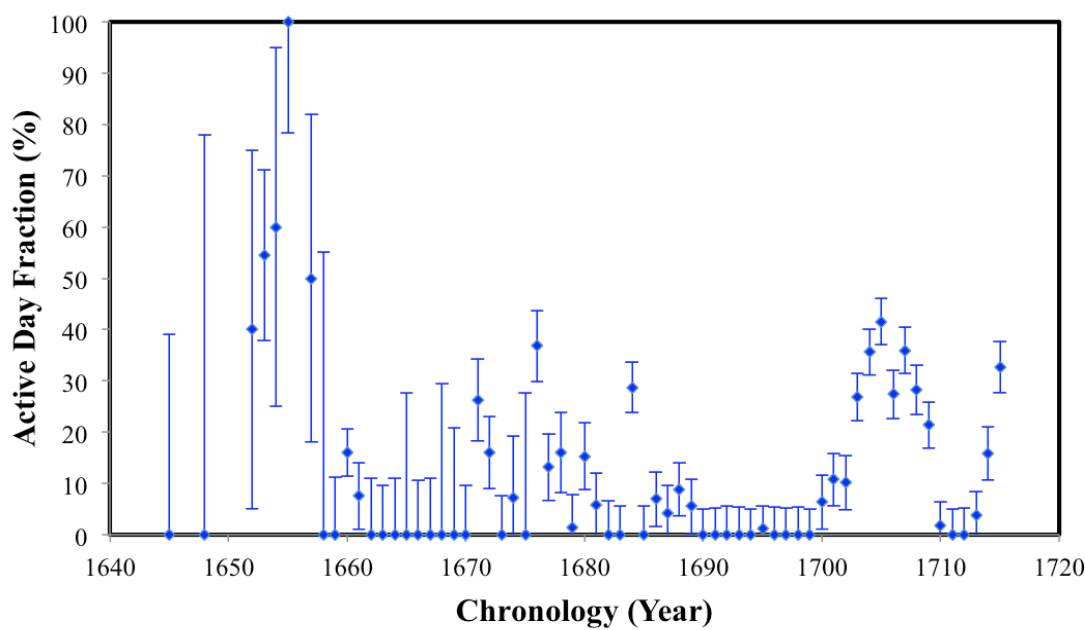

Figure 6: Yearly ADF values in 1645 – 1715 (in percentage). These ADF values are calculated from the sunspot datasets, based on Vaquero *et al*. (2016) and their subsequent updates: Hayakawa *et al*. (2021d) for the Eimmart Observatory, J. H. Müller, Hoffmann, and Wideburg, Hayakawa *et al*. (2021c) for Fogelius, Siverus, Picard, Kircher, and Hook, Carrasco *et al*. (2021b) for Gallet, Neuhäuser *et al*. (2018) for Kirch Family, Ihle, Schultz, Strum, and Hertel, and Hayakawa *et al*. (2024a) for the early sunspot records in 1645 – 1659. See further details in Carrasco *et al*. (2022) and Hayakawa *et al*. (2024a).

## 4. The Maunder Minimum and their Aftermath

Furthermore, the sunspot records allowed us to derive the positions of the reported sunspot groups in the Maunder Minimum. These are particularly important, as the current paradigm comes from Ribes and Nesme-Ribes' seminal study of historical records of the Paris Observatory, which pinpointed the extreme hemispheric asymmetry of the reported sunspot groups in the late Maunder Minimum (Ribes and Nesme-Ribes, 1993).

Notably, to examine independent records to determine whether this trend has been confirmed. This type of study was developed for observations at the Eimmart Observatory, contemporaneous astronomers around Nürnberg, and Johann Christoph Müller. Original records are preserved at the National Library of Russia. Hayakawa *et al*. (2021d, 2021e) examined their records to revise the sunspot group numbers and derive their sunspot positions. Their results located most of the reported sunspot groups in the southern solar hemisphere by 1715, whereas their observations and Johann Christoph Müller's observations from 1716 to 1720 located sunspot groups in both solar hemispheres. This serves as an independent confirmation of the hemispheric asymmetry of the sunspot groups reported in the Maunder Minimum and their relaxation after the Maunder Minimum. Further independent assessments are required to confirm anomalous hemispheric asymmetry.

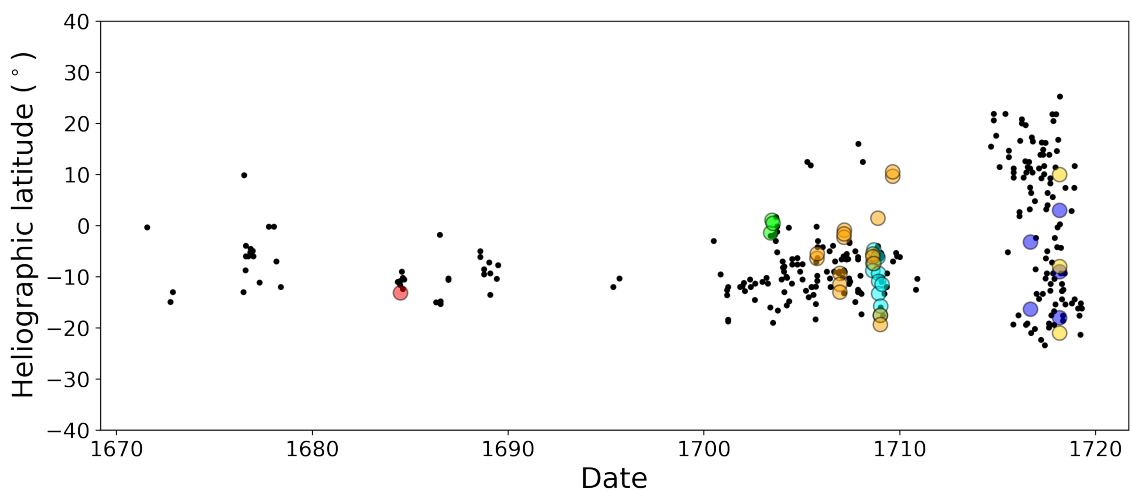

Figure 7: Positions of reported sunspot groups in the archival collections of the Eimmart Observatory and contemporaneous astronomers around Nürnberg (coloured circles) in contrast with the data from La Hire Family in Paris Observatory (Ribes and Nesme-Ribes, 1993) according to Vaquero *et al.*'s (2015b) digitization. This figure is reproduced from Hayakawa *et al.* (2021d).

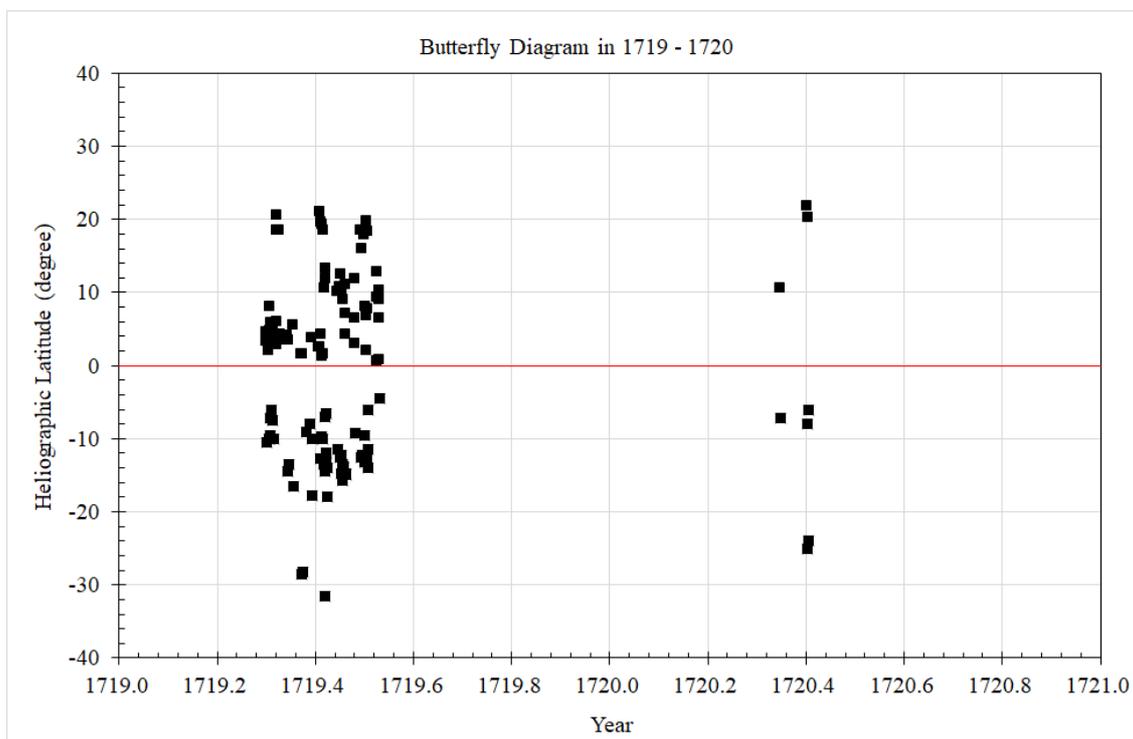

Figure 8: Positions of reported sunspot groups in Johann Christoph Müller's manuscript, as reproduced from Hayakawa *et al*. (2021e).

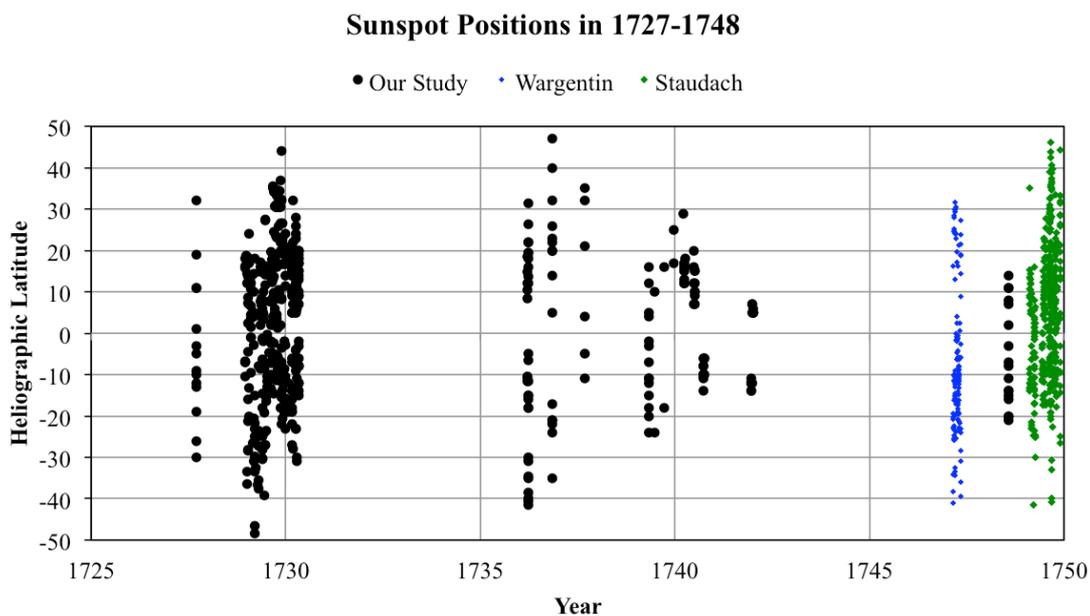

Figure 9: Positions of reported sunspot groups in archival records of contemporaneous sunspot observations in 1727 – 1748, as reproduced from Hayakawa *et al*. (2022).

However, it is unclear whether a similar hemispheric asymmetry was found in the recovery phase from the Maunder Minimum to normal solar cycles, as the sunspot positions in 1721 – 1748 were largely missing owing to poor observational coverage during this period. Hayakawa *et al.* (2022) comprehensively examined the source reports of contemporaneous sunspot records from 1727 to 1748, as long as they were readily accessible. Their results confirmed sunspot groups in both solar hemispheres and their equatorward migration. These trends are consistent with the normal solar cycles in modern sunspot observations, in contrast to the anomalous hemispheric asymmetry of the reported sunspot groups in the Maunder Minimum.

## 5. Summary and Discussion

This presentation briefly reviews recent archival investigations of sunspot data from the 17th to 18th centuries. Kepler's records allowed us to extend reconstructions of sunspot positions back to 1607, locate a probable boundary of solar cycles between 1607 and 1610, and indicate that the first telescopic solar cycle in the 1610s was a normal cycle in terms of duration (Hayakawa *et al.*, 2024b). For the Maunder Minimum, archival investigations have revised considerable datasets, such as the entire records of the early Maunder Minimum, Eimmart Collections, and Hamburg observers, removing contamination and updating active day fractions. These updates directly benefitted the reconstruction of the relative sunspot number and sunspot positions around the Maunder Minimum. The reconstructed sunspot positions emphasized the peculiarity of the Maunder Minimum, even in the 17th and 18th centuries, owing to the extreme hemispheric asymmetry of the reported sunspot groups. Further archival investigations are ongoing to reconstruct past sunspot activity and will improve our understanding of past solar activity.

## Acknowledgments

This research was conducted under the financial supports of JSPS Grant-in-Aids JP20H05643, JP21K13957, and JP22K02956. HH has been part funded by JSPS Overseas Challenge Program for Young Researchers, JSPS Overseas Challenge Program for Young Researchers, the ISEE director's leadership fund for FYs 2021 – 2024, Young Leader Cultivation (YLC) programme of Nagoya University, Tokai Pathways to Global Excellence (Nagoya University) of the Strategic Professional Development Program for Young Researchers (MEXT), TRANSEHA, and the young researcher units for the advancement of new and undeveloped fields, Institute for Advanced Research, Nagoya

University of the Program for Promoting the Enhancement of Research Universities.